\documentclass[doublecol]{epl2}
\usepackage{amssymb}
\usepackage{graphicx}
\usepackage{amsmath}
\usepackage{amsfonts}

\title{Theory for superconductivity in (Tl,K)Fe$_x$Se$_2$ as a doped Mott insulator}

\author{Yi Zhou\inst{1}, Dong-Hui Xu\inst{1}, Fu-Chun Zhang\inst{1,2}, and  Wei-Qiang Chen\inst{2} }

\institute{
  \inst{1} Department of Physics, Zhejiang University, Hangzhou, 310027, P.R. China \\
  \inst{2} Department of Physics and Centre of Theoretical and Computational Physics, The University of Hong Kong, Hong Kong, China
}


\abstract{
Possible superconductivity in recently discovered
(Tl,K)Fe$_x$Se$_2$ compounds is studied from the viewpoint of
doped Mott insulator. The Mott insulating phase is examined to be
preferred in the parent compound at $x=1.5$ due to the presence of
Fe vacancies. Partial filling of vacancies at the Fe-sites
introduces electron carriers and leads to electron doped
superconductivity. By using a two-orbital Hubbard model in the
strong coupling limit, we find that the s-wave pairing is more favorable at small Hund's coupling, 
and d$_{x^2-y^2}$ wave pairing is more favorable at large Hund's coupling. 
}

\begin{document}

\maketitle

Copper-oxide and iron based superconductors are two families with
the highest transition temperatures.\cite{Bednorz,Hosono,Fe-expt1,Fe-expt2,XLChen}.
In copper oxides, the parent compounds are antiferromagnetic (AFM)
Mott insulators. Superconductivity arises when charge carriers are
introduced by chemical doping. In Fe-based materials, the parent
compounds are bad metals with AFM long range order.
Superconductivity arises when the AFM ordering is suppressed by
chemical doping.  A common feature for the superconductivity in
the two families is the nearby AFM states, so that it is generally
believed that the superconductivity is closely related to the
antiferromagnetism in both families. The important difference
between the insulating phase in cuprates and the metallic phase in
Fe-based parent compounds have been thought to distinguish the two
classes of high Tc superconductors. In the theoretical
description, the electron interaction is strong in cuprates and
weak or intermediate in Fe-based compounds.  This has prevented
the development for a unified physical picture for the two
families of high Tc superconductivity. Most theories for Fe-based
superconductivity are based on a weak coupling approach to
consider magnetic
fluctuations as a pairing mechanism.\cite{Fe-theory-WC}
In these theories, the parent compound has a spin density wave
ground state with the gap opening at only part of the Fermi
surface, so that it is a metal.  In the strong coupling theories,
one starts with the assumed insulating parent
state\cite{Si,Kotliar,WQChen} and the theories could only applied to study
the magnetism but have difficulties to explain metallic feature of
the parent state.

Very recently, it has been reported that FeSe-layer compounds
(Tl,K)Fe$_{x}$Se$_2$ are AFM insulators at $1.3<x<1.7$ and become superconductors at $%
1.7<x<1.88 $ with $T_{c}=31$K and $T_{c}^{\mathrm{onset}}=40$K\cite{FangMH}.
Note that there was an early report that K$_{0.8}$Fe$_x$Se$_2$ is
superconducting.\cite{XLChen}  The parent compound of this new family of iron
selenide may be considered to be
 Tl$_{1-y}$K$_{y} $Fe$_{1.5}$Se$_{2}$, where FeSe layers share a similar
structure to FeAs layers in iron arsenic compounds with As$^{3-}$
being replaced by Se$^{2-}$ ions and 1/4 of Fe sites being vacant.
The partial substitution of K by Tl is to stabilize the chemical
component and to prevent the oxidization. This raises an interesting
possibility that Fe-based superconductivity is also a doped Mott
insulator, similar to the cuprates.

In this paper, we propose that the insulating state of
Tl(K)Fe$_{1.5}$Se$_{2}$ is a Mott insulator due to the
Fe-vacancies, which enhance the electron correlation.  The
argument is substantiated by a model calculation involving the
vacancies and the on-site Coulomb and Hund's
coupling.  We use a strong coupling theory of two band model to
show that partial filling of the Fe-vacancies leads to s-wave
superconductivity, which is compatible with the strong disorder in
the system.

We start with the insulating compound Tl(K)Fe$_{1.5}$Se$_{2}$. The
vacancies form a superlattice as suggested in early work and in
the recent transmission electron microscopy on KFe$_{x}$Se$_{2}$
for $1.5\leq x\leq 1.6$.\cite{Haggstrom,LiJQ}
The two possible super-lattice structures are illustrated in Fig.
1, which may be stabilized by the Coulomb repulsion of the
Fe-ions. In this compound, we have all Fe$^{2+}$ or configuration
of Fe-3d$^6$. Local density approximation (LDA) calculations show
a metallic ground state, so that the state is clearly not a band
insulator. The vacancy ordering is a strong evidence that the
insulating phase is not due to disorder effect or the Anderson
localization. We argue that the insulating phase is also difficult
to be explained due to a spin density wave ordering. The gap opened
due to spin density wave ordering is usually at part of the Fermi
surface of the normal state. A full gap at every Fermi point would
require particle-hole symmetry in the electronic structure, which
is not supported by either LDA calculations\cite{LDA}or
angle resolved photoemission spectroscopy(ARPES) data\cite{ARPES}. The
insulating state requires a full gap opening on the Fermi surface,
pointing out the strong electron correlation effect or the Mottness physics
nature of the compound.

\begin{figure}[htbp]
\centerline{\includegraphics[width=0.45\textwidth]{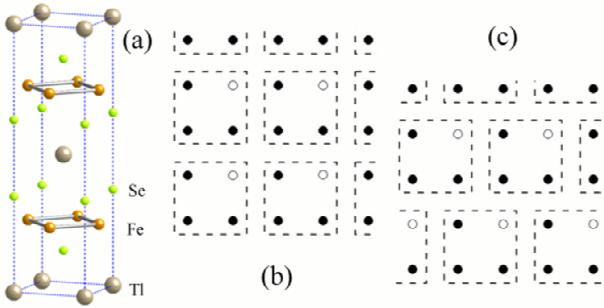}}
\caption{(Color online) TlFe$_{x}$Se$_{2}$ with ThCr$_{2}$Si$_{2}$ type
structure. (a) a unit cell without vacancy; (b) Fe square layer with 1/4
vacancy ordered in square; (c) Fe square layer with 1/4 vacancy ordered in
parallelogram. The filled circles denote Fe atom while the empty circles are
for vacancies. The squares by the dash lines denote the unit cells. Note
that we ignore the up-down distribution of Se atoms which will result in
doubly enlarged unit cells in parallelogram case. }
\end{figure}

Since the parent compounds of Fe-based superconductors in the
absence of Fe-vacancies are metals, the insulating nature of
Tl(K)Fe$_x$Se$_2$ is expected to be related to the Fe-site
vacancies. In the presence of 1/4 Fe-vacancies, the number of
bonds for each Fe-atom to connect with the nearest neighbor (NN)
and the next nearest neighbor (NNN) Fe-atoms is reduced from 8 to
16/3, so that the kinetic energy of Fe-3d electron is
substantially reduced. Note that the electron hopping integrals at
NN and NNN sites are most important kinetic terms in a
tight-binding description for the Fe-based materials. This enhances the
on-site Coulomb repulsion relative to the kinetic energy, hence
the electron correlation. In this scenario, Fe-based compounds are
at the boundary of metal-insulator transition, and may be
"tunable" by introducing superlattice vacancies.  The
superconductivity is induced by chemical doping, which introduces
charge careers in addition to the variation of the Fe-vacancies.
Below we shall use a model to examine the metal-insulator
transition associated with the Fe-vacancies and the
superconductivity in the doped cases.

A generic microscopic model to describe Fe-layer in the system can
be written down in terms of five $3d$ Fe orbitals, which
reads%
\begin{equation}
H=H_{0}+H_{I}.  \label{H}
\end{equation}
$H_{0}$ is a tight-binding Hamiltonian given by%
\begin{equation}
H_{0}=-\sum\limits_{i,\vec{\tau},\alpha \beta \sigma }t_{\vec{\tau}}^{\alpha
\beta }c_{i+\vec{\tau},\alpha \sigma }^{\dag }c_{i\beta \sigma },  \label{H0}
\end{equation}%
where $t_{\vec{\tau}}^{\alpha \beta }$ is the hopping integral between two
sites $i$ and $j=i+\vec{\tau}$ with indices $\alpha ,\beta =1,2,\cdots ,5$
for five 3d orbitals. $H_{I}$ describes the on-site Coulomb interaction,%
\begin{eqnarray}
H_{I} &=&\sum_{i,\alpha }U\hat{n}_{i\alpha \uparrow }\hat{n}_{i\alpha
\downarrow }+\sum_{i,\alpha <\beta }J(\hat{c}_{i\alpha \uparrow }^{\dag }%
\hat{c}_{i\alpha \downarrow }^{\dag }\hat{c}_{i\beta \downarrow }\hat{c}%
_{i\beta \uparrow }+h.c.)  \notag \\
&&+\sum_{i,\alpha <\beta ,\sigma \sigma \prime }(U^{\prime }\hat{n}_{i\alpha
\sigma }\hat{n}_{i\beta \sigma \prime }+J\hat{c}_{i\alpha \sigma }^{\dag
}\hat{c}_{i\beta \sigma \prime }^{\dag }\hat{c}_{i\alpha \sigma \prime }\hat{%
c}_{i\beta \sigma })  \label{HI}
\end{eqnarray}%
where $\hat{n}_{i\alpha \sigma }=\hat{c}_{i\alpha \sigma }^{\dag }\hat{c}%
_{i\alpha \sigma }$, $U$ and $U^{\prime }$ are the intra- and inter-orbital
direct Coulomb repulsions, respectively. $J$ is the Hund's coupling
which satisfies $U=U^{\prime }+2J$ by symmetry.

From quantum chemistry point of view, both the valence of Fe ions
and the buckling of Se ions are similar to those in iron
arsenides, we argue that the low energy electronic states are
mainly Fe-$3d_{xz}$ and $3d_{yz}$ orbitals.  Therefore, we may adopt
a 2-orbital model to study the Mott insulator transition and the
superconductivity at large $U$ limit.  Within this 2-orbital
model, $3d_{x^2-y^2}$ and $3d_{3z^2}$ orbitals are completely occupied
and $3d_{xy}$ orbital is completely empty, so that Fe-3d$^6$
has two electrons, and Fe-3d$^7$ has three electrons or one hole
in the subspace of $3d_{xz}$ and $3d_{yz}$ orbitals. The
simplified 2-orbital model takes similar form of
Eqs.(\ref{H},\ref{H0},\ref{HI}), while $\vec{\tau}$ is for the NN and NNN
bonds only and the orbital indices $\alpha ,\beta =1$ or $2$ are for orbital $%
d_{xz}$ and $d_{yz}$ respectively.  We further set $t_{\hat{x}}^{11}=t_{%
\hat{y}}^{22}=t_{1}$, $t_{\hat{y}}^{11}=t_{\hat{x}}^{22}=t_{2}$, $t_{\hat{x}%
\pm \hat{y}}^{\alpha \alpha }=t_{3}$, and $t_{\hat{x}\pm
\hat{y}}^{12}=\pm t_{4}$ by lattice and orbital symmetry as in
iron pnictides. Within the 2-orbital band model, there are two
electron per Fe-ion in the parent compound Tl(K)Fe$_{1.5}$Se$_2$.
We study metal to Mott insulator transition by using a slave spin
technique\cite{SS}. The main results are shown in Fig.2.  For a
given $J/U$, one sees that the renormalized quasiparticle weight
$Z$ decreases as $U/W$ increases and vanishes at $U=U_{c}$, with W
the bandwidth of the system in the absence of Fe-vacancy.  Our
calculations show that the critical value $U_c$ for the Mott
transition is reduced in the presence of the vacancy, and the
reduction becomes more profound due to the Hund's coupling $J$. The
role of Hund's coupling to the reduction of $U_c$ in multiple-
orbital systems has been studied previously by using slave spin
study on iron pnictides in the absence of Fe-vacancy\cite{SS2}.

\begin{figure}[hpbt]
\includegraphics[width=0.5\textwidth]{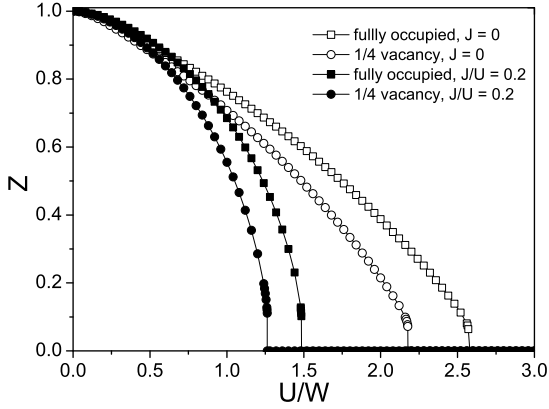}
\caption{Quasiparticle weight $Z$ decreases with increasing $U$ and fixing $%
J/U$. $U_c$ is reduced when vacancy are presented, where the hopping
integrals are chosen to be $t_1=-t,t_2=1.3t,t_3=t_4=-0.85t$, and $W=12t$ is
chosen to be the bandwidth for fully occupied Fe lattice. Vacancies occupy
1/4 Fe sites and forms a square lattice as shown in Fig.1(b).
Note that the results for parallelogram ordered vacancies as shown in Fig.1(c) is quite similar.}
\end{figure}

Electron doping to the parent compound is realized by partial
filling of the Fe-vacancies in the parent compound.
This results in possible superconductivity at low temperatures. Below we
shall examine the pairing interaction of the 2-orbital Hubbard
model from a viewpoint of the doped Mott insulator. In the
superconducting phases, say, $1.7<x<1.88$, vacancy density becomes
lower and a variety of microstructures may coexist. The average
number of electrons per site in the Fe-layer is $4-\frac{3}{x}$ at
the composition Tl(K)Fe$_{x}$Se$_{2}$ as required by the charge balance. At
$x>1.5$, some of the lattice sites have three electrons (or one
hole) within the two-orbital model considered here. These holes
move in the lattice background of Fe-ions with two-hole per site.
The effective interaction between two single hole on
the neighboring sites ($i$,$j$) can be derived by using second
order perturbation theory in the large $(U,\,J)$ limit by
considering the virtual hopping processes\cite{WQChen}, similar to
the super-exchange interaction derived in the single band Hubbard
model. In terms of fermionic representation, this effective
interaction $H_{2}$ can be written in Eq.(\ref{EQ:Heff}) below. We
note that the spin coupling term has been included in study of the
Gossamer superconductivity within the single band Hubbard
model~\cite{Gossamer}.
\begin{eqnarray}
H_{2} &=&-\sum_{ij}\sum_{\alpha \beta \alpha ^{\prime }\beta ^{\prime }}%
\Bigl[A_{\alpha \beta }^{\beta ^{\prime }\alpha ^{\prime }}(ij)\hat{b}%
_{\alpha \beta }^{\dag }(ij)\hat{b}^{\alpha ^{\prime }\beta ^{\prime }}(ij)
\notag \\
&&\phantom{-\sum_{ij}}+\sum_{S_{z}}B_{\alpha \beta }^{\beta ^{\prime }\alpha
^{\prime }}(ij)\hat{T}_{\alpha \beta }^{S_{z}\dag }(ij)\hat{T}%
_{S_{z}}^{\alpha ^{\prime }\beta ^{\prime }}(ij)\Bigr]  \label{EQ:Heff}
\end{eqnarray}%
where $S_{z}=-1,0,1$, and
\begin{eqnarray}
A_{\alpha \beta }^{\beta ^{\prime }\alpha ^{\prime }}(ij) &=&[\frac{%
(-1)^{\beta +\beta ^{\prime }}}{U-J}+\frac{1}{U+J}]t_{ij}^{\alpha
\beta }t_{ji}^{\beta ^{\prime }\alpha ^{\prime }}+\frac{t_{ij}^{\alpha \bar{%
\beta}}t_{ji}^{\bar{\beta ^{\prime }}\alpha ^{\prime }}}{U^{\prime }+J}
\notag \\
B_{\alpha \beta }^{\beta ^{\prime }\alpha ^{\prime }}(ij) &=&\frac{%
(-1)^{\beta +\beta ^{\prime }}}{U^{\prime }-J}t_{ij}^{\alpha \bar{\beta}%
}t_{ji}^{\bar{\beta ^{\prime }}\alpha ^{\prime }},  \label{EQ:ABcoef}
\end{eqnarray}%
where $\bar{\beta}$ indicates the orbital different from $\beta $, and the
first and the second terms in $H_{2}$ are the pairing interaction in the
spin singlet and spin triplet channel, respectively.

We now proceed to discuss the superconducting pairing symmetry.
As discussed above, vacancies are distributed randomly at Fe
layers in superconducting phase and behave similar as non-magnetic
impurities.
The spin singlet s-wave superconductivity is
essentially unaffected by non-magnetic impurities due to
Anderson's theorem\cite{Anderson}, but is strongly affected by
magnetic impurities\cite{BW}. On the other hand, a p-wave
superconductor with spin triplet is very sensitive to both
non-magnetic and magnetic impurities\cite{BW}. This explains why
spin triplet p-wave superconducting state Sr$_{2}$RuO$_{4}$ requires clean sample
and also suggests that the spin triplet pairing is unlikely in
this material.  So we will only consider the spin singlet pairing
below.

Fang et al.'s experiment has suggested that there are several superconducting
transitions in the material\cite{FangMH}.  This result is
consistent with the observed microstructures of the TEM
KFe$_{1.8}$Se$_2$\cite{LiJQ} where both the domains with many ordered
vacancies and the domains with very few randomly distributed vacancies
are found.  As Fe content increases, both the superconductivity
and the total area of the domains with randomly distributed
vacancies increases. As the first step,  we approximate the
disordered system with an average of $4-3/x$ electron per Fe-site
on the Fe-layer. Note that the approximation has an exact limiting
case of no vacancy at $x = 2$, which corresponds to 2.5 electrons
per Fe ion.

By diagonalizing the mean field Hamiltonian of the two orbital
model, one will have two bands \cite{WQChen}, the upper and lower bands with
energy $\epsilon_{\mathbf{k} \pm}$, respectively.  Even in the
case with a large electron concentration, both of the two bands
will still across the Fermi energy and give two electron pockets
around $X$ and $Y$ points and two hole pockets around
$\Gamma$ and $M$ points.  On the other hand, the
ARPES has suggested the disappearance of the hole
pockets\cite{ARPES}.  This discrepancy may be due to the momentum
dependent shift of the Fermi energy in comparing the ARPES and LDA
results\cite{zxshen}. To resolve this discrepancy, here we shall
take a phenomenological approach to assume that the lower bands
are fully occupied and only the upper bands are considered. The
electron carrier concentration is then $\delta = 2 - \frac{3}{x}$.
In the following, we will consider the case with $\delta = 0.3$
which corresponds to $x \approx 1.76$.

\begin{figure}[htbp]
\centerline{\includegraphics[width=0.5\textwidth]{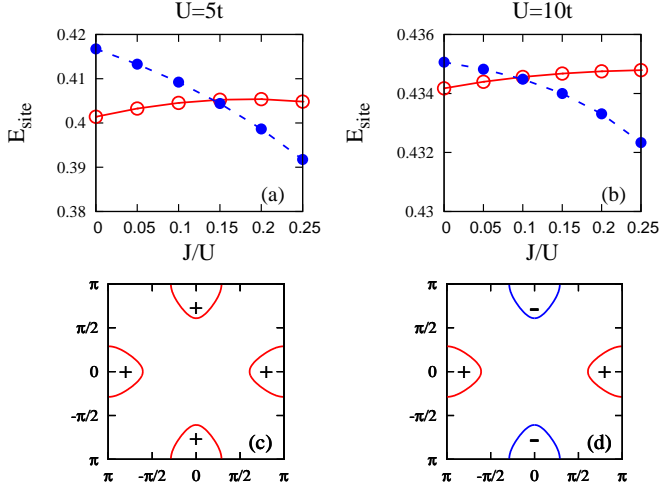}}
\caption[]{\label{fig:energy} (a) and (b), The energy of s-wave-like (red solid line) and d-wave-like (blue dashed line)
  pairing symmetry at $U = 5t$ and $U = 10t$, respectively.  The electron doping is $\delta = 0.3$. (c) and (d), the
  schematic diagram of gap function of $A_{1g}$ and $B_{1g}$ pairing symmetry, respectively.  The former one is like a
  s-wave pairing while the latter one is like a d-wave pairing.  The sign in each pocket indicate the sign of the gap
  function.  The gap function on the Fermi surfaces with same color has same sign.}
\end{figure}

In a random Fe-site vacancy approximation, the symmetry analysis
for superconductivity studied previously for Fe-pnictides can be
applied here, which suggests four possible pairing symmetries in the
even-parity spin-singlet case, i.e. $A_{1g}$, $A_{2g}$, $B_{1g}$
and $B_{2g}$\cite{pairsymmetry}.  So we perform a mean field
calculation similar to that in our previous work\cite{WQChen} to
study the pairing symmetries in FeSe layer compounds.  The main
difference here is that we only consider the pairing between the
electrons in the upper band because the lower band is fully occupied.
Similar with the iron pnictide case, the pairing amplitude in
$A_{2g}$ and $B_{2g}$ channel is very tiny and will not be
discussed further.  The energy of $A_{1g}$ and $B_{1g}$ pairing
symmetries for various $J/U$ with $U = 5t $ and $U = 10t$ are
depicted in Fig.~\ref{fig:energy}.  We see that the
$A_{1g}$ pairing symmetry has lowest energy when $J / U$ is small
while the $B_{1g}$ wins when $J/U$ is large. This result is also
consistent with the qualitative analysis based on the pair hopping
terms\cite{WQChen}.  One difference from the iron pnictide is
that $B_{1g}$ is easier to win in the FeSe layer compounds, which
may be due to the lack of the pairing hopping between electron
pocket and hole pocket, which favors $A_{1g}$ symmetry.

By carefully checking the quasi-particle gap on the Fermi surface,
we find that the gap on each electron pockets are
nodeless and rather isotropic for both $A_{1g}$ and $B_{1g}$ pairing symmetries.
But the gap function of $A_{1g}$
symmetry have same sign on different electron pockets,
while the one of $B_{1g}$ symmetry have different sign on
electron pockets around $X$ and $Y$ point, respectively
 as shown in Fig.~\ref{fig:energy}(c) and (d).  This
indicates that $A_{1g}$ symmetry is like s-wave while $B_{1g}$ symmetry
is d-wave.  But this ``d-wave'' symmetry is nodeless in contrast with the conventional d-wave symmetry.

ARPES experiments show that the superconductor gap is nodeless and almost isotropic.\cite{ARPES} So that the ``s-wave'' pairing state 
appears consistent with ARPES. However, the ``d-wave'' like pairing state can not be ruled out since
it is nodeless and the anisotropy is weak.

The random distribution of Fe vacancies will introduce non-magnetic disorder effect.
Roughly speaking, the vacancies will not affect the ``s-wave'' pairing state drastically, 
but will suppress the superconductivity for other non-s-wave SC states, including the ``d-wave'' like state.
However, the disorder effect in such superconducting systems is a subtle and interesting issue and we will leave
detailed analysis for future study .  We speculate that the non-magnetic impurities act
as what they do in dirty two-band superconductors as studied by Ng using Ginzburg-Landau theory.\cite{TKNg} 
Non-magnetic impurities scatter an electron between $X$ and $Y$ pockets with scattering rate
$\tau_t^{-1}$, where $\tau_t$ is the mean lifetime an electron stays in one pocket.
In the weak scattering regime $\tau_t^{-1} \ll$
the transition temperature $T_c$ or the gap $\Delta$,
the superconductor behave as ``d-wave'' like,
while in the strong scattering regime $\tau_t^{-1} \gg T_c, \Delta$ the superconductor behaves as ``s-wave'' like.



In summary, we studied the Mottness and superconductivity in
(Tl,K)Fe$_x$Se$_2$. By using the slave spin method, we find that
the superlattice vacancies in the parent compound of x=1.5 enhance
the electron correlation and lead to transition to a Mott
insulator. We predict a sizable gap due to the electron correlation in the parent compound ($x=1.5$), which
should be observable in optical measurement~\cite{NLWang}.
Treating Fe vacancies randomly distributed, we find
that a spin-singlet superconductor is likely at such a
doped Mott insulator at $1.7<x<1.88$. The s-wave pairing is more favorable at small Hund's coupling, 
and d$_{x^2-y^2}$ wave pairing is more favorable at large Hund's coupling.
In our theory, the intra-pocket pairing is dominant, the superconductivity is not
relying on the inter-pocket pairing between electron and hole pockets, consisting with the observed superconducting in
only electron pocket systems.  While our study is based on
2-orbital Hubbard model, the essential physics should remain
qualitatively unchanged if more orbitals are included.  It would
be interesting to see the possibility that the insulating state of
the parent compound of x=1.5  become superconducting under a high
pressure, similar to that in layered organic
superconductivity\cite{Gossamer,organic}

We acknowledge valuable discussions with M. H. Fang, P. A. Lee and T. K. Ng.
YZ and DHX are supported by National Basic Research Program of China (Grant No.2011CBA00103),
NSFC (Grant No.11074218) and PCSIRT (Grant No.IRT0754). WQC and FCZ are supported by
RGC grant of HKSAR in part.

{\it Note:} Near the completion of this work, we learned a similar
work by Yu et al. \cite{YuR}, where Mott transition at $x=1.5$
compound is studied.


\begin{thebibliography}{99}


\bibitem{Bednorz}
  \Name{Bednorz J. G. \and Muller K. A.}
  \REVIEW{Z. F. Physik B-Condensed Mater}{64}{1986}{189-193}.


\bibitem{Hosono}
  \Name{Kamihara Y., \textit{et al.}}
  \REVIEW{J. Am. Chem. Soc}{130}{2008}{3296}.


\bibitem{Fe-expt1}
  \Name{Chen X. H., \textit{et al.}}
  \REVIEW{Nature}{453}{2008}{761};
  \Name{Chen G. F., \textit{et al.}}
  \REVIEW{Phys. Rev. Lett.}{100}{2008}{247002};
  \Name{Wen H.H., \textit{et al.}}
  \REVIEW{EuroPhys. Lett.}{82}{2008}{17009};
  \Name{Wang C., \textit{et al.}}
  \REVIEW{Europhys. Lett.}{83}{2008}{67006};
  \Name{Ren Z. A., \textit{et al.}}
  \REVIEW{Europhys. Lett.}{83}{2008}{17002}.


\bibitem{Fe-expt2}
  \Name{Rotter M., \textit{et al.}}
  \REVIEW{Phys. Rev. B}{78}{2008}{020503(R)};
  \Name{Hsu F. C., \textit{et al.}}
  \REVIEW{PNAS}{105}{2008}{14262};
  \Name{Fang M. H., \textit{et al.}}
  \REVIEW{Phys. Rev. B}{78}{2008}{224503};
  \Name{Tapp J. H., \textit{et al.}}
  \REVIEW{Phys. Rev. B}{78}{2008}{060505(R)};
  \Name{Wang X. C., \textit{et al.}}
  \REVIEW{Solid State Commun.}{148}{2008}{538}.


\bibitem{XLChen}
  \Name{Guo J. G., \textit{et al.}}
  \REVIEW{Phys. Rev. B}{82}{2010}{180520(R)}.


\bibitem{Fe-theory-WC}
  \Name{Mazin I. I., \textit{et al.}}
  \REVIEW{Phys. Rev. Lett.}{101}{2008}{057003};
  \Name{Dai X., \textit{et al.}}
  \REVIEW{Phys. Rev. Lett.}{101}{2008}{057008};
  \Name{Raghu S., \textit{et al.}}
  \REVIEW{Phys. Rev. B}{77}{2008}{220503(R)};
  \Name{Lee P. A. \and Wen X. G.}
  \REVIEW{Phys. Rev. B}{78}{2008}{144517};
  \Name{Wang F., \textit{et al.}}
  \REVIEW{Phys. Rev. Lett.}{102}{2009}{047005};
  \Name{Yao Z. J., \textit{et al.}}
  \REVIEW{New J. Phys.}{11}{2009}{025009};
  \Name{Chubukov A. V., \textit{et al.}}
  \REVIEW{Phys. Rev. B}{78}{2008}{134512};
  \Name{Kuroki K., \textit{et al.}}
  \REVIEW{Phys. Rev. Lett.}{101}{2008}{087004};
  \Name{Cvetkovic V. \and Tesanovic Z.}
  \REVIEW{EuroPhys. Lett.}{85}{2009}{37002}.


\bibitem{Si}
	\Name{Si Q. M. \and Abrahams E.}
	\REVIEW{Phys. Rev. Lett.}{101}{2008}{076401}.


\bibitem{Kotliar}
	\Name{Haule K., \textit{et al.}}
	\REVIEW{Phys. Rev. Lett.}{100}{2008}{226402}.


\bibitem{WQChen}
	\Name{Chen W. Q., \textit{et al.}}
	\REVIEW{Phys. Rev. Lett.}{102}{2009}{047006}.


\bibitem{FangMH}
	\Name{Fang M. H., \textit{et al.}}
	\REVIEW{EuroPhys. Lett.}{94}{2010}{27009}.


\bibitem{Haggstrom}
	\Name{H\"{a}ggstr\"{o}m L., \textit{et al.}}
	\REVIEW{J. Solid State Chem.}{63}{1986}{401}.


\bibitem{LiJQ}
	\Name{Wang Z., \textit{et al.}}
	\REVIEW{e-preprint}{}{2011}{arXiv:1101.2059}.


\bibitem{LDA}
	\Name{Yan X. W., \textit{et al.}}
	\REVIEW{Phys. Rev. Lett.}{106}{2011}{087005};
	\Name{Cao C. \and Dai J. H.}
	\REVIEW{e-preprint}{}{2011}{arXiv:1101.0533}.


\bibitem{ARPES}
	\Name{Zhang Y., \textit{et al.}}
	\REVIEW{Nature Materials}{10}{2011}{273};
	\Name{Mou D. X., \textit{et al.}}
	\REVIEW{Phys. Rev. Lett.}{106}{2011}{107001};
	\Name{Wang X. P., \textit{et al.}}
	\REVIEW{EuroPhys. Lett.}{93}{2011}{57001}.


\bibitem{SS}
	\Name{de'Medici L., \textit{et al.}}
	\REVIEW{Phys. Rev. B}{72}{2005}{205124}.


\bibitem{SS2}
	\Name{Yu R. \and Si. Q. M.}
	\REVIEW{e-preprint}{}{2011}{arXiv:1006.2337}.


\bibitem{Gossamer}
	\Name{Zhang F. C.}
	\REVIEW{Phys. Rev. Lett.}{90}{2003}{207002};
	\Name{Laughlin R.}
	\REVIEW{e-preprint}{}{2002}{cond-mat/0209269};
	\Name{Gan J. Y., \textit{et al.}}
  \REVIEW{Phys. Rev. Lett.}{94}{2005}{067005}.


\bibitem{Anderson}
	\Name{Anderson P. W.}
	\REVIEW{Phys. Rev. Lett.}{3}{1959}{325}.


\bibitem{BW}
	\Name{Balian R. \and Werthamer N. R.}
	\REVIEW{Phys. Rev.}{131}{1963}{1560}.


\bibitem{zxshen}
	\Name{Yi M. {\it et al.}}
	\REVIEW{Phys. Rev. B}{80}{2009}{024515}.


\bibitem{pairsymmetry}
	\Name{Zhou Y., \textit{et al.}}
	\REVIEW{Phys. Rev. B}{78}{2008}{064514};
	\Name{Wang Z. H., \textit{et al.}}
	\REVIEW{e-preprint}{}{2008}{arXiv:0805.0736};
	\Name{Wan Y. \and Wang Q. H.}
	\REVIEW{Europhys. Lett.}{85}{2009}{57007};
	\Name{Shi J. R., \textit{et al.}}
	\REVIEW{e-preprint}{}{2008}{arXiv:0806.0259}.


\bibitem{organic}
	\Name{Lefebvre S. {\it et al.}}
	\REVIEW{Phys. Rev. Lett.}{85}{2000}{5420};
	\Name{Sasaki T. {\it et al.}}
	\REVIEW{Phys. Rev. B}{65}{2002}{060505(R)};
	\Name{Muller J. {\it et al.}}
	\REVIEW{Phys. Rev. B}{65}{2002}{144521}.

\bibitem{TKNg}
	\Name{Ng T. K.}
	\REVIEW{Phys. Rev. Lett.}{103}{2009}{236402}.


\bibitem{NLWang} We note that a tiny gap was reported in optical measurement on K$_{0.8}$Fe$_{2-x}$Se$_2$, see
Z. G. Chen {\it et. al.}, arXiv:1101.0572. It remains to be seen if the best parent sample
Tl(K)Fe$_{1.5}$Se$_2$ will show a sizable Mott gap in optical measurement.



\bibitem{YuR}
	\Name{Yu R., \textit{et al.}}
	\REVIEW{e-preprint}{}{2011}{arXiv:1101.3307}.

\end{thebibliography}
\end{document}